\documentclass[preprintnumbers,showpacs,floats,twocolumn,prd,aps]{revtex4}
\usepackage{amssymb,amsmath,epsfig}
\usepackage{bm}

\begin{document}

\setlength{\pdfpagewidth}{8.5in}

\setlength{\pdfpageheight}{11in}

\title{Self-gravitating Interferometry and Intrinsic Decoherence}
\author{Cisco Gooding}\email{cgooding@physics.ubc.ca}
\author{William G. Unruh}\email{unruh@physics.ubc.ca}
\affiliation{
  Department of Physics and Astronomy,
  University of British Columbia\\
  Vancouver, British Columbia, V6T 1Z1 Canada
}
\date{\today}

\begin{abstract}
To investigate the possibility that intrinsic gravitational decoherence can be theoretically demonstrated within canonical quantum gravity, we develop a model of a self-gravitating interferometer. We search for evidence in the resulting interference pattern that would indicate coherence is fundamentally limited due to general relativistic effects. To eliminate the occurence of gravitational waves, we work in spherical symmetry, and construct the ``beam'' of the interferometer out of WKB states for an infinitesimally thin shell of matter. For internal consistency, we encode information about the beam optics within the dynamics of the shell itself, by arranging an ideal fluid on the surface of the shell with an equation of state that enforces beam-splitting and reflections. We then determine sufficient conditions for (interferometric) coherence to be fully present even after general relativistic corrections are introduced, test whether or not they can be satisfied, and remark on the implications of the results.
\end{abstract}

\maketitle

\section{Introduction}

Despite not having a complete theory of quantum gravity, it is becoming more and more important to understand systems for which both quantum and general relativistic effects are important (see \cite{RuoBerchera13}-\cite{Lamine06}, for instance, and references therein). Indeed, it is the study of such systems that helps clarify the clash between quantum theory (QT) and general relativity (GR), in hope that we may find guidance towards a resolution of the many technical and conceptual problems one faces when attempting to unify these two pillars of physics.

The purpose of this work is to explore an ambiguity the results from taking both QT and GR seriously: quantum superpositions of different matter states are associated with different spacetime geometries, and hence different definitions of time evolution; how, then, can we use a single time-evolution operator to evolve superpositions of distinct spacetimes? Since the $1980$s, Roger Penrose has been arguing that this ambiguity results in an instability, and that in turn this instability leads to a type of ``decay'' that reduces the system to a state with a single well-defined geometry \cite{Penrose86}, \cite{Penrose}. It is not clear from Penrose's work, however, whether some sort of ``collapse'' occurs, or whether there is simply a form of ``intrinsic'' decoherence that removes phase correlations between states associated with sufficiently different geometries. In this paper, we consider the latter, and discuss whether or not a direct application of both QT and GR is enough to demonstrate the existence of this new type of ``intrinsic'' decoherence.

By ``intrinsic'' decoherence, we mean a decoherence effect that arises solely out of the internal behaviour of an isolated system, and not due to its interaction with the external world. For example, if we use a buckyball in a double-slit experiment, and prepare one of the slits to excite internal degrees of freedom of the buckyball, then the internal degrees of freedom carry ``which-way'' information and decohere the center-of-mass degree of freedom \cite{Banaszek13}. More generally, if a system carries an internal clock and is in a superposition of states corresponding to two paths that have different proper times associated with them, then again the internal clock read at the interference screen could provide which-way information, and decohere the center-of-mass \cite{PZCB13}-\cite{ZCPRB12}.

Whereas the decoherence produced by entangling internal degrees of freedom to a center-of-mass coordinate could be considered ``third-party'' decoherence \cite{Stamp06}, what we are concerned with here is whether or not there is something about gravity itself that could lead to such intrinsic decoherence. Penrose's intuition says yes: the path a mass takes alters the associated spacetime and especially the flow of time. Since the quantum phase is determined by the flow of time, the phase evolution is also altered by which path the mass takes. When one tries to interfere the two paths, these ``random'' phases (because it is impossible to uniquely map one spacetime onto another) cause decoherence. It is this \textit{gravitational} intrinsic decoherence that we explore here.

There are several proposals in the literature for a mechanism to describe gravitational intrinsic decoherence \cite{Diosi}-\cite{GambiniPullin06}, and other proposals for intrinsic decoherence mechanisms that are merely inspired by tension between QT and GR \cite{Milburn91}-\cite{Percival95}. Many of these approaches incorporate alterations to known physics, such as adding stochastic \cite{Percival95} or nonlinear \cite{Weinberg89I}-\cite{Penrose98} terms to the Schr\"{o}dinger equation, in order to achieve the desired decoherence effect. Such alterations are often ad hoc, and have historically faced difficulties maintaining consistency with experimental constraints; nonlinear additions to the Schr\"{o}dinger equation, for instance, have been shown under a wide range of conditions to lead to either superluminal signal propagation or to communication between different branches of the wavefunction \cite{Polchinsky91}. While it may be possible to obtain a sensible theory that allows communication between different wavefunction branches \cite{Stamp}, it remains to be seen whether a consistent interpretation results from this alteration. Instead, we take Penrose's initial arguments at face value, and entertain the possibility that a consistent combination of QT and GR can explain (gravitational) intrinsic decoherence without any assumptions about new physics. 

Now, it is well-known that gravitational waves can carry away information from a system in a manner analogous to standard decoherence \cite{Wang06}, \cite{Hu14}. Penrose's suggestion is independent of such standard gravitational-wave-induced decoherence, so to distinguish between the two we will work in spherical symmetry. The restriction to spherical symmetry is not only a technical simplification, but avoids the occurrence of gravitational waves altogether. 

Because this exploration requires both QT and GR, we will naturally be faced with some serious difficulties, which we will have to either overcome, sidestep, or ignore \cite{DeWitt67}. For instance, we will avoid issues with the factor-ordering ambiguity by working in the WKB regime (as in \cite{Visser}), and we will avoid issues with perturbative non-renormalizability by working in minisuperspace (i.e. enforcing spherical symmetry) and employing a reduced phase space approximation (as in \cite{Kuchar71}). It is still unclear what the exact connection is between the reduced phase space approximation, obtained by solving the GR constraints classically and then quantizing the reduced theory, and the standard Dirac quantization, obtained by quantizing the theory in the full kinematical Hilbert space and then enforcing the constraints at the quantum level. Following Hawking's path integral approach to quantum gravity \cite{Hawking78}, Halliwell has made some progress elucidating the connection between reduced phase space minisuperspace quantization and the standard Dirac approach in special cases \cite{Halliwell88}, but in general the connection is not well understood. Nonetheless, the limit we will work in has a rich structure, and in this paper we will explore whether or not it has a rich enough structure to contain evidence of intrinsic decoherence caused by gravity. 

Since we aim to test whether or not gravity places a fundamental limit on the coherence of quantum systems, we develop a model of a self-gravitating interferometer. Interferometers are ideal for studying coherence, because interference is a key feature of coherent systems. We describe how the same interferometer would behave in the absence of gravity, and then we investigate the consequences of general relativistic corrections to this behaviour. In an interferometric setting, the intrinsic decoherence we seek to understand manifests itself as a phase-scrambling along different interferometer arms (for a general discussion see \cite{Aharonov90}), which in this case is attributed to gravity. According to Penrose, we should expect that the (interferometric) coherence should decay as the arm-length increases indefinitely, since this would correspond to a superposition of arbitrarily different spacetimes. We focus on the possibility that no collapse occurs, so we will simply analyze the interference pattern and search for departures from non-gravitational behaviour that indicate coherence loss. Conceptually, we are testing the idea that when one forms superpositions of geometries in the interferometer, the nature of time in GR leads at the quantum level to an imprint of which-way information, which is accompanied by a loss of fringe visibility \cite{Englert96}.

Still, an objection may be raised that if one describes the interferometer as a closed quantum system without tracing out over any physical degrees of freedom, then QT implies that coherence must prevail, regardless of whether the system is general relativistic. This objection was raised by Banks, Susskind, and Peskin \cite{BanksSusskind84} in the context of black hole evaporation, but it was later pointed out not only that the arguments in \cite{BanksSusskind84} were inconclusive, but that we have reason to support the possibility that pure states can effectively evolve into mixed states in black hole systems \cite{UnruhWald95}.

The more radical idea entertained here is that one might find pure states evolving to mixed states in gravitational systems without horizons. In general, this ``dissipationless'' type of decoherence has been explored to some degree \cite{Stamp88}-\cite{Gangopadhyay01}, but even the fact that it is possible has not been widely appreciated. Nonetheless, one can observe that the thermal character of acceleration radiation is approximately present even without the involvement of Rindler horizons (for recent analyses see \cite{Rovelli12}, \cite{Langlois}), and by the equivalence principle one might expect to find a gravitational analog of this thermal behaviour. This means, then, that one might expect that gravity generates an intrinsic form of entropy, even in systems without the horizon structure that one usually associates with entropy in black hole thermodynamics.

With this in mind, we will construct our interferometer, theoretically, out of a self-gravitating, spherically symmetric, infinitesimally thin shell of matter. The interferometer ``optics'' are encoded internally, by adding tangential pressure to the fluid that lives on the surface of the shell. The resulting model is reminiscent of an idea Einstein first proposed in $1939$ \cite{EinsteinCluster}, but in our case, the tangential pressure satisfies an equation of state that produces a beam-splitter and perfect reflectors. The fluid is ideal, in the sense that one obtains a perfect-fluid stress-energy tensor, if one projects the full four-dimensional spacetime stress-energy tensor onto the three-dimensional history of the shell. This approach ensures that the interferometric setup is manifestly invariant under coordinate transformations.

The configuration we construct resembles that of a Michelson interferometer in optics. Thus, we will send initial states at a beam-splitter, at which point the transmitted and reflected components travel in opposite directions until they encounter ``mirrors.'' The components will then reflect, travel back towards each other, and encounter the splitter once more. There will be two possible outputs, corresponding to final transmission and final reflection, which are comprised of different combinations of the initially split wave components.  

What we mean by (interferometric) coherence, in this system, is the sustained phase relationships between different wave components that can allow us, for instance, to completely cancel either of the final outputs. In other words, if we are unable to obtain complete constructive or destructive interference in our interferometer (as predicted by Penrose), we can conclude that coherence is being limited in the system. The goal of our current investigation is to determine whether or not general relativistic effects could demonstrably produce such a limitation.

\section{Theory of Self-Gravitating Spherical Shells}

\subsection{Action Principle}

The perfect-fluid shell model we develop is a generalization of the dust shell model used by Kraus and Wilczek in their attempt to calculate self-interaction corrections to standard Hawking radiation \cite{KrausWilczek}-\cite{KrausCharged}. Generalizing the Kraus and Wilczek approach to include the required pressure effects is not without complications, even in the classical theory. In contrast to the approach to thin-shells pioneered by Israel that involves stitching two spacetimes together along the shell's history \cite{Israel}, the starting point for our theoretical considerations is an action that is composed of a gravitational part given by the Einstein-Hilbert action, plus some action for the shell that we can initially leave unspecified, written (in natural units) as
\begin{equation}
I = \frac{1}{16\pi}\int{d^4x\,\sqrt{-g^{(4)}} \hspace{3pt} \mathcal{R}^{(4)}} + I_{shell}.
\end{equation}
The superscripts on the metric determinant $g$ and the Ricci scalar $\mathcal{R}$ indicate that these quantities are constructed from components of the full spacetime metric $g_{\mu\nu}$, with $\mu,\nu \in \{0,1,2,3\}$. 

We will express the metric in ADM form \cite{ADM}, which in spherical symmetry is given by
\begin{equation}
g_{\mu\nu}dx^{\mu}dx^{\nu}=-N^2 dt^2 + L^2\left(dr+N^r dt\right)^2+R^2d\Omega^2,
\label{eq:Metric}
\end{equation}
where $N$ is the lapse function, $N^r$ is the radial component of the shift vector, and $L^2$ and $R^2$ are the only nontrivial components of the spatial metric.

The angular variables are taken to be the polar angle $\theta$ and the azimuthal angle $\phi$, such that the angular metric takes the form $d\Omega^2=d\theta^2+\sin^2{\theta} d\phi^2$.

The shell action studied by Kraus and Wilczek takes the form
\begin{equation}
I_{dust}=-m \int{d\lambda\, \sqrt{-g_{\mu\nu}\frac{d x^\mu}{d\lambda}\frac{d x^\nu}{d\lambda}}},
\end{equation}
with $m$ being the rest-mass of the shell and all metric quantities evaluated on the shell history. The arbitrary parameter $\lambda$ can be chosen to coincide with the coordinate time $t$, which simplifies the integrand for the shell action.

To describe a more general fluid than dust, we need a more general action. There are well-established variational principles for regular perfect fluids in GR \cite{Schutz}, but the authors are unaware of any satisfactory variational principles for the perfect fluid shells we wish to describe. The stress-energy tensor for a perfect fluid with density $\sigma$ and pressure $p$ is given by
\begin{equation}
S^{ab}=\sigma u^a u^b + p ( \gamma^{ab}+u^a u^b ),
\label{eq:FluidTensor}
\end{equation}
where $u^a$ are the components of the fluid proper velocity in coordinates that cover the fluid history. For our purposes, the geometry along the fluid history of our shell is described by an induced metric $\gamma_{ab}dy^a dy^b = -d\tau^2+\hat{R}^2d\Omega^2$, with $\tau$ being the shell proper time. This induced metric obeys the relation
\begin{equation}
\gamma_{ab} = e_a^{\mu}e_b^{\nu}g_{\mu\nu},
\end{equation}
with the introduction of projectors onto the shell history given by
\begin{equation}
e_a^{\mu}=\frac{\partial x^{\mu} }{\partial y^a}=u^\mu \delta_a^\tau + \delta_\Omega^\mu \delta_a^\Omega.
\end{equation}
Here and elsewhere, the repeated $\Omega$ denotes a sum over angular coordinates. These projectors allow us to express the full spacetime stress-energy tensor of our perfect fluid shell as
\begin{equation}
T^{\mu\nu}=S^{ab}e_a^\mu e_b^\nu \delta (\chi),
\label{eq:FullTensor}
\end{equation}
where we have introduced a Gaussian normal coordinate $\chi$ in the direction of the outward-pointing space-like unit normal $\bm{\xi}$, with the shell location defined by $\chi=0$.

We want to obtain an action, expressed in terms of the full spacetime quantities, that yields the tensor (\ref{eq:FluidTensor}) in the intrinsic coordinates of the shell history. To convert derivative expressions from the intrinsic coordinates to the ADM coordinates given in equation (\ref{eq:Metric}), we can write infinitesimal changes in $r$ and $t$ as
\begin{equation}
dt = u^t d\tau + \xi^t d\chi, \hspace{5pt}
dr = u^r d\tau + \xi^r d\chi.
\label{eq:CoordDiffs}
\end{equation}
Taking advantage of the fact that $\bm{\xi}$ satisfies $u^\mu \xi_\mu=0$ and $\xi^\mu \xi_\mu = 1$, and suppressing the (vanishing) angular components for brevity, the outward normal can be written as
\begin{equation}
\xi_\alpha=\sqrt{g_{tr}^2-g_{tt}g_{rr}}\left(-u^r,u^t\right)=N^2 L^2\left(-u^r,u^t\right).
\end{equation}

For radial integration within an ADM slice, one has $dt=0$, and in this case we can solve for $\frac{dr}{d\chi}$ in equation (\ref{eq:CoordDiffs}) to obtain \cite{DeltaConversion}
\begin{equation}
\frac{dr}{d\chi}=\xi^r-\frac{u^r}{u^t}\xi^t.
\end{equation}
Also, since $u^\mu=(\partial t / \partial \tau)(1,\dot{X},0,0)$, the $4$-velocity normalization $u^\mu u_\mu=-1$ (evaluated on the shell) implies
\begin{equation}
\left(\frac{\partial t}{\partial \tau}\right)^2=(u^t)^2=\left(N^2-L^2(N^r+\dot{X})^2\right)^{-1}.
\label{eq:Normalization}
\end{equation}

This allows conversion of the delta function appearing in our expression (\ref{eq:FullTensor}) for the full spacetime stress-energy tensor:
\begin{equation}
\delta (\chi) = \frac{dr}{d\chi}\delta (r-X)=\frac{\sqrt{N^2-L^2(N^r+\dot{X})^2}}{N L}\delta(r-X).
\label{eq:Delta}
\end{equation}

Using equation (\ref{eq:FullTensor}), we find that our stress-energy tensor takes the form
\begin{equation}
T^{\mu\nu}=\left(\sigma u^\mu u^\nu + p \hspace{1pt} g^{\Omega\Omega}\delta_\Omega^\mu \delta_\Omega^\nu \right) \delta(\chi),
\label{eq:FullFluidTensor}
\end{equation}
where the repeated $\Omega$ indices denote a single sum over angular coordinates. In expression (\ref{eq:FullFluidTensor}), the ``tangential'' nature of the pressure is manifest, since the projection of this tensor onto the space-like normal $\bm{\xi}$ clearly vanishes.

The action we seek, then, yields (\ref{eq:FullFluidTensor}) upon taking variations with respect to the metric, in accordance with the definition
\begin{equation}
\delta I = \frac{1}{2}\int{d^4x\,\sqrt{-^4g}T^{\mu\nu}\delta g_{\mu\nu}}.
\label{eq:StressDefinition}
\end{equation}
We are especially interested in the contribution from the tangential pressure, which takes the form
\begin{equation}
\delta I_p = 8\pi\int{dt\,dr\,N L \delta\left(\chi\right)p R \delta R}.
\end{equation}
By inspection, we find that the action
\begin{equation}
I_{shell}=- \int{d\lambda \,\sqrt{-g_{\mu\nu}\frac{d x^\mu}{d\lambda}\frac{d x^\nu}{d\lambda}}} M(R),
\label{eq:ShellAction}
\end{equation}
with all quantities evaluated on the shell history, yields the appropriate stress-energy tensor: the relevant variational derivative of (\ref{eq:ShellAction}) with respect to the metric is
\begin{equation}
\delta I_{shell,p} = -\int{dt\,dr\,N L \delta\left(\chi\right)M'(R) \delta R},
\end{equation}
from which it follows that one has the pressure identification $p=-M'(R)/8\pi R$, along with the usual density identification $\sigma=M(R)/4\pi R^2$. We will use the freedom in choosing the function $M(R)$ to parametrize an equation of state that relates the density and pressure of our fluid. It should be noted that $R$ is not a coordinate, but a metric component that serves as a measure of the shell's internal energy.

The action (\ref{eq:ShellAction}) is reparametrization-invariant, as well as invariant under general (spherically symmetric) coordinate transformations, even with the inclusion of an $R$-dependent `mass'. As mentioned above, this is because $R$, when evaluated on the shell, is nothing more than the reduced area of the shell, and this area is independent of coordinate choices.

\subsection{Hamiltonianization}

Following the canonical formalism \cite{ADM}, one can perform a Legendre transformation $\mathcal{H}=P\dot{X}-\mathcal{L}$, for the shell variables. Here $\mathcal{L}$ is the Lagrangian defined by (\ref{eq:ShellAction}), subject to the condition that the shell history is parametrized by $t$. One then finds
\begin{equation}
\mathcal{L}=-\int{dr\,\sqrt{N^2-L^2(N^r+\dot{X})^2}M(R)\delta(r-X)},
\label{eq:ShellL}
\end{equation}
and it follows that the momentum conjugate to the shell position $X$ for the unreduced problem is given by
\begin{equation}
P=\frac{\partial \mathcal{L}}{\partial \dot{X}}=\int{dr\,\frac{L^2 (N^r+\dot{X}) M(R)}{\sqrt{N^2-L^2(N^r+\dot{X})^2}}\delta(r-X)}.
\label{eq:ShellMomentum}
\end{equation}
Explicitly, we can determine the Hamiltonian $\mathcal{H}$ to be
\begin{equation}
\mathcal{H}=P\dot{X}-\mathcal{L}=\int{dr \,\left(NH_0^s+ N^rH_r^s\right)},
\end{equation}
with the definitions
\begin{eqnarray}
H_0^s &=& \sqrt{L^{-2}P^2+M(R)^2}\delta(r-X), \nonumber \\ 
H_r^s &=& -P\delta(r-X).
\end{eqnarray}

Similarly, we can Hamiltonianize the gravitational action, and express the total action as
\begin{equation}
I = \int{dt\, P\dot{X}}+\int{dt\,dr\,\left(\pi_R \dot{R}+\pi_L \dot{L}-NH_0-N^r H_r\right)},
\end{equation}
for $H_0=H_0^s+H_0^G$ and $H_r=H_r^s+H_r^G$, such that
\begin{eqnarray}
H_0^G &=& \frac{L\pi_L^2}{2R^2}-\frac{\pi_L\pi_R}{R}+\left(\frac{R R'}{L}\right)'-\frac{(R')^2}{2L}-\frac{L}{2}, \nonumber\\
H_r^G &=& R'\pi_R-L\pi_L'.
\label{eq:GravConstraints}
\end{eqnarray}

\subsection{Equations of Motion}

Once in Hamiltonian form, the equations of motion for the system are obtained by varying the action with respect to the variables $N$, $N^r$, $\pi_L$, $\pi_R$, $L$, and $R$. Explicitly, these variations (respectively) lead to
\begin{eqnarray}
\label{eq:GravEOM}
H_0 &=& 0, \nonumber\\
H_r &=& 0, \nonumber\\
\dot{L} &=& \frac{N}{R}\left(\frac{L\pi_L}{R}-\pi_R\right)+\left(N^r L\right)', \nonumber\\
\dot{R} &=& -\frac{N\pi_L}{R}+N^r R', \\  
\dot{\pi_L} &=& \frac{N}{2}\left(1-\frac{\pi_L^2}{R^2}-\frac{(R')^2}{L^2}\right)-\frac{N' R R'}{L^2} \nonumber\\
&&+ N^r \pi_L'+\frac{N P^2 \delta(r-X)}{L^2\sqrt{P^2+L^2 M^2}}, \nonumber\\
\dot{\pi_R} &=& \frac{N\pi_L}{R^2}\left(\frac{L\pi_L}{R}-\pi_R\right)-N\left(\frac{R'}{L}\right)'-\left(\frac{N'R}{L}\right)' \nonumber\\
&&+ \left(N^r \pi_R\right)' - \frac{NM\frac{dM}{dR}\delta(r-X)}{\sqrt{L^{-2} P^2+M^2}}. \nonumber
\end{eqnarray}
The first two equations are the Hamiltonian and momentum constraints, whereas the next four are the dynamical equations of motion for the gravitational variables.

For the shell variables, the equation of motion for $X$ can be easily obtained by varying the action with respect to $P$, or simply by solving equation (\ref{eq:ShellMomentum}) for $\dot{X}$. The result is
\begin{eqnarray}
\dot{X} &=& \int{dr\,\left(\frac{NP}{L\sqrt{P^2+L^2 M^2}}-N^r\right)\delta(r-X)} \nonumber\\
&=& \frac{\hat{N}P}{\hat{L}\sqrt{P^2+\hat{L}^2 \hat{M}^2}}-\hat{N^r},
\label{eq:XEOM}
\end{eqnarray}
with hats indicating that one evaluates the quantities at $r=X$. 

The equation of motion for $P$ is more subtle, since a standard variation of the action with respect to $X$ is formally ambiguous, as noted in \cite{FriedmanLuoko}. The ambiguity arises because one must evaluate quantities on the shell ($L'$, $(N^r)'$, $N'$ and $R'$) that are (possibly) discontinuous at $r=X$:
\begin{equation}
\dot{P}=\left(N^r P - N\sqrt{L^{-2}P^2+M^2}\right)_{shell}'.
\label{ambiguous}
\end{equation}

However, it has been demonstrated in \cite{Menotti} that this ambiguity can be removed by requiring consistency with the constraints and the gravitational equations of motion (\ref{eq:GravEOM}), at least for the case of a dust shell. The argument described in \cite{Menotti} shows that one must average the discontinuous quantities when interpreting the equation of motion for the shell momentum, and similar reasoning leads to the same conclusion for the arbitrary perfect fluid shell described here. One then has the equation of motion
\begin{equation}
\dot{P}=\bar{(N^r)'}P-\frac{\bar{N'}}{\hat{L}}\sqrt{P^2+\hat{L}^2 \hat{M}^2}+\frac{\hat{N}\left(P^2\bar{L'}-\hat{L}^3 \hat{M} \bar{M'}\right)}{\hat{L}^2\sqrt{P^2+\hat{L}^2 \hat{M}^2}},
\label{eq:Pdot}
\end{equation}
with the average taken over $(N^r)'$ in the first term of the right-hand-side, and the last term containing the factor $\bar{M'}$ defined as $\bar{M'}=\hat{\frac{dM}{dR}}\bar{R'}$.

Let us briefly sketch the argument that leads to this result. To start, we take the time derivative of the (integrated and rearranged) momentum constraint:
\begin{equation}
\dot{P}=-\Delta\pi_L\frac{d}{dt}\left(\hat{L}\right)-\hat{L}\frac{d}{dt}\left(\Delta\pi_L\right).
\label{eq:constraintderivative}
\end{equation}
Then, by continuity of $\dot{L}$, we have
\begin{equation}
\frac{d}{dt}L(X)=L'(X\pm\epsilon)\dot{X}+\dot{L}(X\pm\epsilon)=\bar{L'}\dot{X}+\bar{\dot{L}}.
\end{equation}
Averaging the equation of motion for $L$, noting that $\frac{d}{dt}(\Delta\pi_L)=\Delta(\pi_L')\dot{X}+\Delta(\dot{\pi_L})$, and calculating $\Delta(\dot{\pi_L})$ from the equation of motion for $\pi_L$, we obtain
\begin{equation}
\dot{P}=\mathcal{\dot{P}}+\Phi,
\end{equation}
with $\mathcal{\dot{P}}$ representing the right side of equation (\ref{eq:Pdot}), and $\Phi$ defined such that
\begin{eqnarray}
\Phi=&-&P\frac{\hat{N}}{\hat{R}\hat{L}}\bar{\pi_R}+\frac{\hat{N}\Delta R'\bar{R'}}{\hat{L}}+\Delta N'\frac{\bar{R'}\hat{R}}{\hat{L}} \nonumber\\
&-&\hat{L}\Delta\pi_L'(\hat{N^r}+\dot{X})+\frac{\hat{N}\hat{M}\hat{\frac{dM}{dR}}\bar{R'}}{\sqrt{\hat{L}^{-2}P^2+\hat{M}^2}}.
\end{eqnarray}
To then demonstrate that $\Phi$ vanishes, one needs to take the jump of the momentum constraint across the shell to obtain $\hat{L}\Delta\pi_L'=\bar{R'}\Delta\pi_R+\bar{\pi_R}\Delta R'$, then integrate the equation of motion for $\pi_R$ across the shell, and use the result, combined with the fact that the delta contribution to $\dot{\pi_R}$ is given by $-\dot{X}(\Delta\pi_R)\delta(r-X)$ \cite{FriedmanLuoko}.

\subsection{Phase Space Reduction}

We now seek a description of the system in terms of only the shell coordinate $X$ and a conjugate momentum $P_c$. Note that it is not necessarily true that $P_c$ will coincide with the conjugate momentum $P$ for the unreduced problem, as will become clear in what follows. 

To proceed with the Hamiltonian reduction, we will make use of the Liouville form $\mathcal{F}$ and the symplectic form $\Omega$, which on the full phase space (denoted by $\Gamma$) can be written as
\begin{equation}
\mathcal{F}=P_c \bm{\delta} X + \int{dr\left(\pi_L \bm{\delta} L +\pi_R \bm{\delta} R\right)}
\end{equation}
and
\begin{equation}
\Omega = \bm{\delta} P_c \wedge \bm{\delta} X + \int{dr \left(\bm{\delta} \pi_L \wedge \bm{\delta} L + \bm{\delta} \pi_R \wedge \bm{\delta} R \right)}, 
\end{equation}
respectively, with $\bm{\delta}$ denoting an exterior derivative in the associated functional space (see \cite{FriedmanLuoko} for more details). The reduced phase space $\bar{\Gamma}$ is defined as the set of equivalence classes in $\Gamma$ under changes of coordinates, and each (permissible) choice of coordinates defines a hypersurface $\bar{H}\subseteq \Gamma$ that is transversal to the orbits generated by coordinate transformations; this ensures that there exists an isomorphism between $\bar{\Gamma}$ and the representative hypersurface $\bar{H}$. 

At this point we can determine the symplectic form $\bar{\Omega}$ induced on $\bar{H}$ as follows: first, consider the pullback of $\mathcal{F}$ to $\bar{H}$; this yields a quantity which we denote by $\mathcal{F}_{\bar{H}}$. Then, the symplectic form $\Omega_{\bar{H}}$ on the representative hypersurface $\bar{H}$ (corresponding to $\bar{\Omega}$) takes the form 
\begin{equation}
\Omega_{\bar{H}}=\bm{\delta} \mathcal{F}_{\bar{H}}.
\end{equation}
This quantity defines the canonical structure of the reduced phase space.

To explicitly determine the (nonlocal) contribution of the gravitational variables to the dynamics on the reduced phase space, we can solve the GR constraints for the gravitational momenta, insert the solutions into the Liouville form on the full phase space, and perform the integration to express the gravitational contribution solely in terms of the (local) shell variables. Away from the shell, take the following linear combination of the constraints:
\begin{equation}
-\frac{R'}{L}H_0-\frac{\pi_L}{RL}H_r=\mathcal{M}',
\end{equation}
for
\begin{equation}
\mathcal{M}(r)=\frac{\pi_L^2}{2R}+\frac{R}{2}-\frac{R(R')^2}{2L^2}.
\end{equation}
The quantity $\mathcal{M}(r)$ corresponds to the ADM mass $H$ when evaluated outside of the shell, and vanishes inside the shell. This enables us to solve for the gravitational momenta $\pi_L$, $\pi_R$ away from the shell. The result is
\begin{equation}
\pi_L=\pm R\sqrt{\left(\frac{R'}{L}\right)^2-1+\frac{2\mathcal{M}}{R}},\hspace{8pt}\pi_R=\frac{L}{R'}\pi_L'.
\label{eq:GravMomenta}
\end{equation}

One then makes a coordinate choice, to pick out a representative hypersurface $\bar{H}$. The coordinates we will use resemble the flat-slice coordinates \{$L=1$, $R=r$\} described in \cite{KrausPainleveGullstrand} (also known as Painlev\'e-Gullstrand coordinates), though we will have a deformation region $X-\epsilon<r<X$ explicitly included, in order to both satisfy the constraints and yield a continuous spatial metric. The deformation region is related to a jump in $R'$ across the shell. This can be seen by first integrating the Hamiltonian and momentum constraints across the shell. Doing so yields, respectively,
\begin{equation}
\Delta R'=-\frac{\hat{V}}{\hat{R}}, \hspace{8pt} \Delta\pi_L=-\frac{P}{\hat{L}},
\label{eq:Jumps}
\end{equation}
where $V=\sqrt{P^2+M^2L^2}$ and $\Delta$ indicates the jump of a quantity across the shell. We therefore take
\begin{equation}
L=1, \hspace{8pt} R(r,t)=r-\frac{\epsilon}{X}\hat{V}f\left(\frac{X-r}{\epsilon}\right),
\label{eq:Coords}
\end{equation}
for a function $f$ having support in the interval $(0,1)$ with the property $f'(x)\rightarrow 1$ as $x\searrow0$. Outside of the deformation region, these coincide with flat-slice coordinates. For concreteness, let us suppose $f(x)$ takes the form 
\begin{equation}
f(x)=x e^{-x^2/(1-x^2)}
\end{equation}
for all $x \in (0,1)$.

In what follows, it will be useful to note that $\hat{M}=M(\hat{R})=M(X)$, and that now $P$ is considered to be a function of $X$ and $H$, as a consequence of the gravitational constraints. We can implicitly determine this function by inserting the gravitational momentum solutions away from the shell given by equation (\ref{eq:GravMomenta}) into the jump equations (\ref{eq:Jumps}) and squaring. We are then left with
\begin{equation}
H=\hat{V}+\frac{\hat{M}^2}{2X}-P\sqrt{\frac{2H}{X}}.
\label{eq:DeterminesP}
\end{equation}

With this coordinate choice, the only gravitational contribution to the Liouville form comes from the $\pi_R$ term, and only from within the deformation region. Keeping in mind that we only care about terms that remain nonzero in the $\epsilon\rightarrow0$ limit, we have, in the deformation region,
\begin{equation}
\pi_R = \frac{X R''}{\sqrt{\left(R'\right)^2-1}}+\mathcal{O}(1),
\end{equation}
since $R=X+\mathcal{O}(\epsilon)$ and $R''=\mathcal{O}\left(\epsilon^{-1}\right)$. One can also note that $\bm{\delta} R = \left(1-R'\right)\bm{\delta} X+\mathcal{O}(\epsilon)$, and express the gravitational contribution to the Liouville form as
\begin{equation}
\int_{X-\epsilon}^{X}{dr\, \pi_R \bm{\delta} R}=X\bm{\delta} X\int_{X-\epsilon}^{X}{dr\,\frac{R'' \left(1-R'\right)}{\sqrt{\left(R'\right)^2-1}}}+\mathcal{O}(\epsilon).
\end{equation}
To evaluate this integral, one can change the integration variable from $r$ to $v=R'$:
\begin{equation}
\int_{X-\epsilon}^{X}{dr\, \pi_R \bm{\delta} R}=X\bm{\delta} X\int_{1}^{R_-'}{dv\,\frac{\left(1-v\right)}{\sqrt{v^2-1}}}+\mathcal{O}(\epsilon),
\end{equation}
with $R_-'$ being $R'$ evaluated just inside the shell. The integration is then straightforward, and after applying (\ref{eq:DeterminesP}) and making some rearrangements one arrives at
\begin{equation}
X\bm{\delta} X \left[-\frac{P}{X}-\sqrt{\frac{2 H}{X}}+\ln{\left(1+\sqrt{\frac{2H}{X}}+\frac{\hat{V}+P}{X}\right)}\right],
\end{equation}
plus terms that vanish as $\epsilon\rightarrow 0$. This completes the calculation of $\mathcal{F}_{\bar{H}}$, the pullback of the full Liouville form $\mathcal{F}$ to $\bar{H}$:
\begin{equation}
\mathcal{F}_{\bar{H}}=P_c \bm{\delta} X,
\end{equation}
with the reduced canonical momentum evidently given by
\begin{equation}
P_c = -\sqrt{2HX}+X\ln{\left(1+\sqrt{\frac{2H}{X}}+\frac{\hat{V}+P}{X}\right)}.
\end{equation}
This result agrees with \cite{FriedmanLuoko} in the limit of a dust shell ($\hat{M}'=0$).

To connect this with the expression derived by Kraus and Wilczek, we need only apply the expression (\ref{eq:DeterminesP}) to the argument of the logarithm, which leads to
\begin{equation}
P_c = -\sqrt{2HX}-X\ln{\left(\frac{X+\hat{V}-P-\sqrt{2HX}}{X}\right)}.
\label{eq:Pc}
\end{equation}
This form of the reduced momentum coincides with \cite{KrausWilczek} in the dust-shell limit.

\subsection{Boundary Terms}

To obtain a well-defined variational principle for the reduced problem, we must be careful with boundary terms, as first noted in \cite{RT1974} and \cite{Unruh1976}. In \cite{KrausWilczek}, it is observed that for asymptotically-flat spacetimes, we simply need to subtract the ADM mass (denoted suggestively by $H$) from our reduced Lagrangian. Specifically, as mentioned in \cite{FriedmanLuoko}, a nonzero boundary variation results from integrating by parts the term $\int{dtdrN^rL(\delta \pi_L)'}$, which is part of the momentum constraint. The only contribution comes from infinity, and in this case we have $N^r \rightarrow N\sqrt{2H/r}$, $N\rightarrow 1$, and 
\begin{equation}
\delta (\pi_L)\rightarrow \delta(\sqrt{2Hr}) = \sqrt{\frac{r}{2H}}\delta H,
\end{equation}
so the variation of the boundary term is cancelled if we add to the action the term
\begin{equation}
I_{bdry}=-\int{dt\, H}.
\end{equation}

Including the boundary term to the action defined by $\mathcal{F}_{\bar{H}}$, one obtains the reduced action
\begin{equation}
I_{reduced}=\int{dt\, \left(P_c \dot{X} - H\right)},
\label{eq:ReducedAction}
\end{equation}
with the reduced momentum given by (\ref{eq:Pc}). From the form of the reduced action (\ref{eq:ReducedAction}), we can see that the ADM mass is the reduced Hamiltonian, defined implicitly by (\ref{eq:Pc}) and (\ref{eq:DeterminesP}).

Since (\ref{eq:DeterminesP}) has more than one solution $P=P(X,H)$, our conjugate momentum $P_c$ in turn becomes a multi-valued function of $X$ and $H$, as one expects from a theory that allows the degree of freedom to either increase or decrease. Explicitly, $P$ is given by
\begin{eqnarray}
&P = \frac{1}{1-\frac{2H}{X}}\left(\sqrt{\frac{2H}{X}}\left(H-\frac{\hat{M}^2}{2X}\right)\right)& \nonumber\\
&\pm \frac{1}{1-\frac{2H}{X}}\left(\sqrt{\left(H-\frac{\hat{M}^2}{2X}\right)^2-\hat{M}^2\left(1-\frac{2H}{X}\right)}\right),&
\end{eqnarray}
while the combination $\hat{V}-P$ that appears in the reduced momentum (\ref{eq:Pc}) is 
\begin{equation}
\hat{V}-P=\frac{H-\frac{\hat{M}^2}{2X}\mp \sqrt{\left(H-\frac{\hat{M}^2}{2X}\right)^2-\hat{M}^2\left(1-\frac{2H}{X}\right)}}{1+\sqrt{\frac{2H}{X}}}.
\end{equation}

\subsection{Constructing Classical Spacetime}

Suppose one can find a solution $X(t)$ to the classical equations of motion for the reduced system (\ref{eq:ReducedAction}). Then, the gravitational constraints and equations of motion (\ref{eq:GravEOM}) can be solved to determine all the metric components $g_{\mu\nu}$. Therefore, from the reduced system solution $X(t)$ one can construct the classical spacetime structure, as we will now demonstrate.

By inserting the gravitational momenta solutions (\ref{eq:GravMomenta}) into the gravitational equations of motion (\ref{eq:GravEOM}), one can obtain the lapse function $N$ and the radial shift component $N^r$ that correspond to our coordinate choice (\ref{eq:Coords}).

Outside of the shell, one finds the familiar Schwarzschild structure, in flat-slice coordinates. The lapse function is constant, and unity if we want a time coordinate that increases towards the future, while the radial shift is given by
\begin{equation}
N^r(r\geq X)= \pm \sqrt{\frac{2H}{r}}.
\end{equation} 
The $\pm$ here indicates two possible time-slicings, though we will often take the upper sign (this means that the gravitational momenta solution (\ref{eq:GravMomenta}) should take the upper sign as well, to ensure $N\rightarrow 1$ as $r\rightarrow \infty$).

Along with the expression (\ref{eq:DeterminesP}) for $P$ in terms of $X$ and $H$, we now have enough information to determine the classical path $X(t)$, since $H$ is constant along such paths. Specifically, the equation of motion for $X$ becomes

\begin{equation}
\dot{X}=\frac{P}{\sqrt{P^2+\hat{M}^2}}-\sqrt{\frac{2H}{X}},
\end{equation}
which leads to the expression
\begin{eqnarray}
\frac{dt}{dX} &=& \frac{\sqrt{2H X}}{X-2H}  \\
&\pm& \frac{H-\frac{\hat{M}^2}{2X}}{\left(1-\frac{2H}{X}\right)\sqrt{\left(H-\frac{\hat{M}^2}{2X}\right)^2-\hat{M}^2\left(1-\frac{2H}{X}\right)}}. \nonumber
\end{eqnarray}
Therefore, finding the classical path $X(t)$ has been reduced to quadrature and inversion.

With the classical path known, one can also calculate the classical action, as done for the case of dust in \cite{KrausWilczek}:
\begin{equation}
S(t,X(t))=S(0,X(0))+\int_0^t{d\tilde{t}\left[P_c(\tilde{t})\dot{X}(\tilde{t})-H\right]},
\end{equation}
with
\begin{equation}
P_c(0)=\frac{\partial S}{\partial X}(0,X(0)).
\end{equation}
Unlike the (massless) dust case, however, our classical path $X(t)$ is not a null geodesic of the flat-slice metric
\begin{equation}
ds^2=-dt^2+\left(dr+\sqrt{\frac{2H}{r}}dt\right)^2,
\end{equation}
and so we cannot so easily determine explicit expressions for our shell trajectories.

\section{Interferometry}

\subsection{Equation of State Determination}

Up until this point, the function $M(X)$ has been left unspecified, though we have established the identifications $\sigma=M(X)/4\pi X^2$ for the density and $p=-M'(X)/8\pi X$ for the pressure. We would like to exploit this freedom for the purposes of interferometry. To maintain internal consistency, there should be a relationship $p=p(\sigma)$, which represents an equation of state for our fluid shell. The function $M(X)$ parametrizes this relationship, though not every choice of $M(X)$ yields a consistent (let alone physical) equation of state.

\begin{figure}[fig:Interferometer]
\includegraphics[width=3.5in]{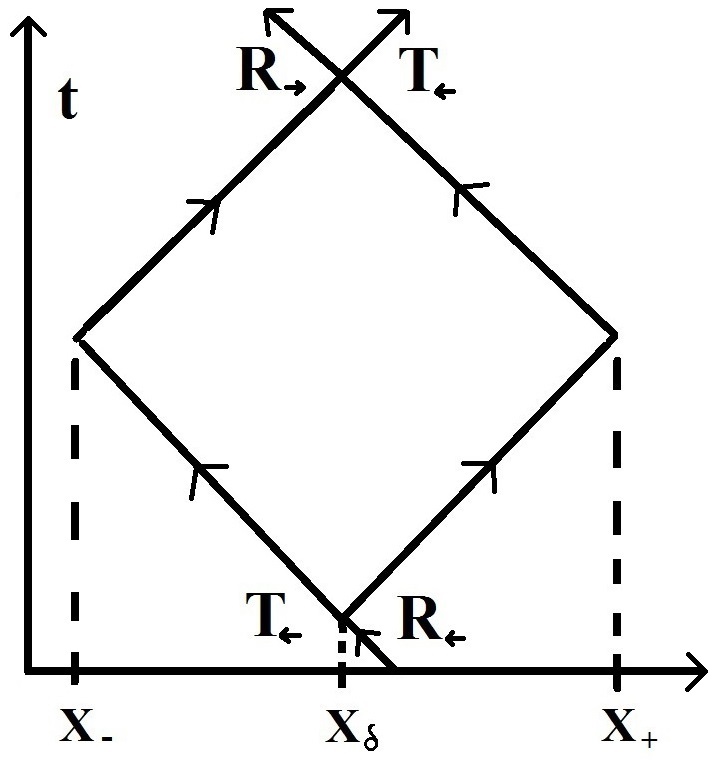}
\caption{Schematic representation of the splitting, reflecting, and recombining that occur in our shell interferometer. $R$ and $T$ are effective reflection and transmission coefficients for the beam-splitter.}
\label{fig:Interferometer}
\end{figure}

The interferometric setup resembles that of Michelson, except we only have one spatial dimension to work with, since our system is spherically symmetric. Still, we would like the equation of state to produce two `reflectors' - one to reflect the shell outward if it gets too small, and one to reflect the shell inward if it gets too large. Also, we would like the equivalent of a `half-silvered mirror' to be in between the two reflectors, to act as a beam-splitter. This is depicted schematically in Figure~\ref{fig:Interferometer}, with $X_\pm$ being the shell radii that correspond to the reflectors, and $X_\delta$ the radius corresponding to the splitter. Accordingly, our equation of state $p=p(\sigma)$ must have a large positive peak for some large density, a large negative peak for some small density, and an intermediate peak for some intermediate density.

It would be convenient to use delta functions for these purposes, but due to the conversion between $\delta(\sigma-\sigma_0)$ and $\delta(X-X_0)$ and the resulting appearance of products of delta functions, this possibility seems problematic. Therefore, we have been considering the simplest alternative one could think of: rectangular barriers. These can be described with the use of step functions, which we will define such that $\Theta(x<0)=0$ and $\Theta(x>0)=1$.

The equation of state, then, takes the form
\begin{eqnarray}
p &=& p_1 \left( \Theta(\sigma-\sigma_1)-\Theta(\sigma-\sigma_2) \right) \nonumber \\
 && + p_2 \left( \Theta(\sigma-\sigma_3)-\Theta(\sigma-\sigma_4) \right) \nonumber \\
 && + p_3 \left( \Theta(\sigma-\sigma_5)-\Theta(\sigma-\sigma_6) \right),
 \label{eq:EOS}
\end{eqnarray} 
with $\sigma_{i+1}>\sigma_{i}$, $p_1<0$, and $p_2, p_3 > 0$. We may as well take $p_1=-p_3$, since both of these peaks serve the same purpose of reflecting, but we will not yet impose this condition. 

\begin{figure}
  \centerline{\epsfig{file=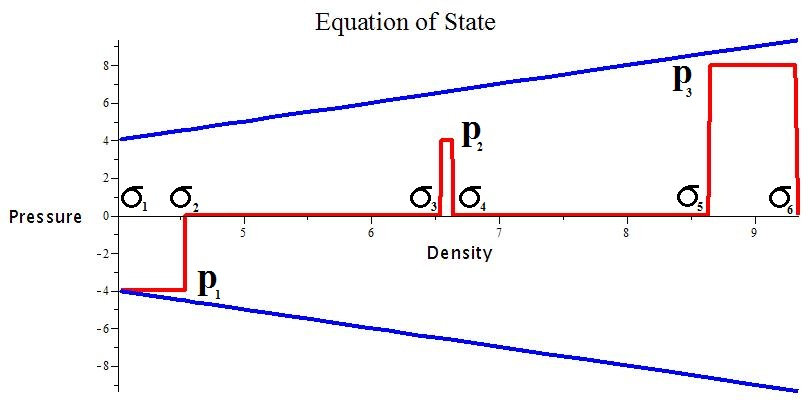, width=3.5in}}
  \caption{
    A sample equation of state represented by (\ref{eq:EOS}).
  }
  \label{fig:EOS}
\end{figure}
Figure~\ref{fig:EOS} illustrates the desired step function peaks, to enable our system to operate as an interferometer.

One would now like to find the function $M(X)$ that parametrizes the equation of state (\ref{eq:EOS}). If we could express (\ref{eq:EOS}) as $p = \sum_i{\tilde{p}_i \Theta (X-X_i)}$, then the identification $p=-M'(X)/8\pi X$ would imply
\begin{equation}
\label{eq:MassFunction}
M(X)=M_0+4\pi\sum_i{\tilde{p}_i\left(X_i^2-X^2\right)\Theta(X-X_i)},
\end{equation}
which would yield a density given by
\begin{equation}
\sigma = \frac{M_0}{4\pi X^2}+\sum_i{\tilde{p}_i\left(\frac{X_i^2}{X^2}-1\right)\Theta(X-X_i)}.
\end{equation}
The problem with this possibility is that, in general, it isn't necessarily true that $\Theta(X-X_i)$ produces the same (reversed) ordering as $\Theta(\sigma-\sigma_i)$, given that $\sigma_i=M(X_i)/4\pi X_i^2$. This problem can be avoided by making sure that the density $\sigma$ is a monotonically decreasing function of $X$. This leads to the condition
\begin{equation}
\frac{M_0}{4\pi} \geq -\sum_{i}\tilde{p}_i X_i^2 \Theta(X-X_i).
\end{equation}

To understand what this means in terms of the pressure peaks in our equation of state (\ref{eq:EOS}), we first note that if $\sigma$ monotonically decreases in $X$, then step functions can be converted by $\Theta(\sigma-\sigma_i) = 1-\Theta(X-X_i)$. This allows us to conclude that $\tilde{p_2}=-\tilde{p_1}=p_1$, $\tilde{p_4}=-\tilde{p_3}=p_2$, and $\tilde{p_6}=-\tilde{p_5}=p_3$. Then, one finds that monotonicity is maintained as long as
\begin{eqnarray}
\frac{M_0}{4\pi} > max\{ p_3 X_{5,6}^2, p_3 X_{5,6}^2+p_2 X_{3,4}^2, \nonumber \\
  p_3 X_{5,6}^2+ p_2 X_{3,4}^2-p_1 X_2^2 \},
\end{eqnarray}
where the notation $X_{i,j}^2=X_i^2-X_j^2$ was introduced, for brevity.

Since an equation of state (\ref{eq:EOS}) is described by the pressure as a function of density, one should translate the conditions for monotonicity in terms of the step locations $\{\sigma_i\}$ and the step amplitudes $\{p_i\}$. To convert between the $\{X_i\}$ and the $\{\sigma_i\}$, one can use the relations
\begin{eqnarray}
X_6^2 &=&\frac{M_0}{4\pi \sigma_6}, \nonumber\\
X_5^2&=&\frac{M_0}{4\pi \sigma_6}\frac{(\sigma_6+p_3)}{(\sigma_5+p_3)}, \nonumber\\
X_4^2&=&\frac{M_0}{4\pi \sigma_6}\frac{(\sigma_6+p_3)}{(\sigma_5+p_3)}\frac{\sigma_5}{\sigma_4}, \nonumber\\
X_3^2&=&\frac{M_0}{4\pi  \sigma_6}\frac{(\sigma_6+p_3)}{(\sigma_5+p_3)}\frac{\sigma_5}{\sigma_4}\frac{(\sigma_4+p_2)}{(\sigma_3+p_2)}, \nonumber\\
X_2^2&=&\frac{M_0}{4\pi  \sigma_6}\frac{(\sigma_6+p_3)}{(\sigma_5+p_3)}\frac{\sigma_5}{\sigma_4}\frac{(\sigma_4+p_2)}{(\sigma_3+p_2)}\frac{\sigma_3}{\sigma_2}, \nonumber\\
X_1^2&=&\frac{M_0}{4\pi  \sigma_6}\frac{(\sigma_6+p_3)}{(\sigma_5+p_3)}\frac{\sigma_5}{\sigma_4}\frac{(\sigma_4+p_2)}{(\sigma_3+p_2)}\frac{\sigma_3}{\sigma_2}\frac{(\sigma_2+p_1)}{(\sigma_1+p_1)}.
\label{eq:Convert}
\end{eqnarray}
With these expressions, one can write the monotonicity conditions in the much simpler form 
\begin{equation}
\{\sigma_5 > 0,\sigma_3 > 0, \sigma_1+p_1 > 0 \}.
\end{equation}
Thus, as long as we keep the density $\sigma$ positive, it will be monotonic in $X$ provided $\sigma_1+p_1 > 0$.

\subsection{Flat Spacetime Limit}
To determine whether or not gravity produces some form of decoherence in our interferometer, let us first clarify the manner in which coherence manifests itself in the absence of gravity. In this case spacetime is flat, and along the arms of the interferometer defined by (\ref{eq:MassFunction}) the shell behaves as a free particle.

\begin{figure}
  \centerline{\epsfig{file=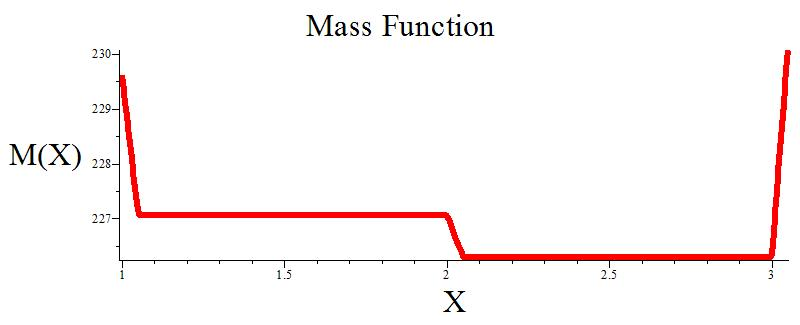, width=3.5in}}
  \caption{A sample mass function $\hat{M}$ is plotted with respect to the shell radius $X$. The approximate step function near $X=2$ serves as a beam-splitter, and the steep quadratic sides correspond to the inner and outer reflectors of the interferometer.}
  \label{fig:MassFunction}
\end{figure}
As evident from Figure~\ref{fig:MassFunction}, the mass of the ``free" shell is different on each interferometer arm. Let us call the inner mass $M_-$, and the outer mass $M_+$, such that $M_->M_+$. For simplicity, suppose the reflectors are perfect, which for this system means that the quadratic walls of the mass function are large and steep. Similarly, let the quadratic beam-splitter interval be approximated by a step function, to ensure that only constant mass function basis states need to be used in the quantum analysis.

Further, let us treat each element of the interferometer separately, in a similar manner to that which is done in optical systems. The initial state will first encounter the splitter, at which point each incoming mode will transform into a reflected mode with a factor $R_\leftarrow$ and a transmitted mode with a factor $T_\leftarrow$ (subscripts are used here because the reflection/transmission coefficients depend on the direction the incoming state encounters the splitter from).

The split initial state components will then perfectly reflect off of the outer/inner reflectors, and travel back towards one another to the beam-splitter. Upon recombination there will be further splitting of the components coming from each direction of the splitter, which produces two outputs (one going in each direction from the splitter) that are themselves composed of two parts; it is the interference between these two parts of each output that we are interested in.

Let us now describe the process in detail. For the purposes of this paper, we will restrict our attention to a single-mode input, since the multi-mode analysis is more involved and will be presented elsewhere \cite{GU14}. We will approximate the single-mode input by an ingoing WKB state:
\begin{equation}
\Psi_0= \frac{e^{i\int{dX P_{-+}}}}{\sqrt{|\partial E/\partial P_{-+}}|}\equiv \psi_{-+},
\end{equation}
where the first set of plus/minuses of the reduced momentum $P$ indicating outgoing/ingoing, and the second set indicating evaluations of $P$ as $X$ approaches $X_\delta$ from above/below (we have dropped the subscript $c$ on the reduced momentum here and for the rest of the paper, for brevity). We will define the integration such that the lower bound in $X$ is $X_\delta$.

Treating the first splitting on its own, let us consider the wavefunction
\begin{displaymath}
   \Psi = \left\{
     \begin{array}{lr}
       \psi_{-+}+R_{\leftarrow}\psi_{++} & : X > X_\delta \\
       T_{\leftarrow}\psi_{--} & : X < X_\delta
     \end{array}
   \right.
\end{displaymath}
The (classical) flat spacetime Hamiltonian satisfies $H=\sqrt{P^2+\hat{M}^2}$, which in the nonrelativistic limit yields $H\approx \hat{M}+P^2/2\hat{M}$. Applying wavefunction continuity at $X_\delta$, and integrating the nonrelativistic Schr\"{o}dinger equation
\begin{equation}
i\frac{\partial \psi}{\partial t}= \hat{M}\psi- \frac{1}{2}\frac{\partial}{\partial X}\left( \frac{1}{\hat{M}}\frac{\partial \psi}{\partial X} \right)
\end{equation}
across $X_\delta$, one can obtain the reflection and transmission amplitudes $R_\leftarrow$ and $T_\leftarrow$. The equations take a simpler form after transforming to the variables $\bar{R}_\leftarrow$ and $\bar{T}_\leftarrow$, which are defined by
\begin{equation}
\bar{R}_{\leftarrow} \equiv \sqrt{\left|\frac{\frac{\partial E}{\partial P_{-+}}}{\frac{\partial E}{\partial P_{++}}}\right|} R_\leftarrow, \hspace{6pt}
\bar{T}_{\leftarrow} \equiv \sqrt{\left|\frac{\frac{\partial E}{\partial P_{-+}}}{\frac{\partial E}{\partial P_{--}}}\right|} T_\leftarrow.
\end{equation}
One can then easily solve for the new variables:
\begin{equation}
\bar{R}_{\leftarrow}=\frac{M_- P_{-+}-M_+ P_{--}}{M_+ P_{--}-M_- P_{++}}, \hspace{6pt} \bar{T}_\leftarrow=\frac{M_- (P_{++}-P_{-+})}{M_- P_{++}-M_+ P_{--}}.
\end{equation}
For convenience, we can also derive the reflection and transmission amplitudes from the left, which are found to be
\begin{equation}
\bar{R}_{\rightarrow}=\frac{M_- P_{++}-M_+ P_{+-}}{M_+ P_{--}-M_- P_{++}}, \hspace{6pt} \bar{T}_\rightarrow=\frac{M_+ (P_{+-}-P_{--})}{M_- P_{++}-M_+ P_{--}},
\end{equation}
using the similar definitions
\begin{equation}
\bar{R}_{\rightarrow} \equiv \sqrt{\left|\frac{\frac{\partial E}{\partial P_{+-}}}{\frac{\partial E}{\partial P_{--}}}\right|} R_\rightarrow, \hspace{6pt}
\bar{T}_{\rightarrow} \equiv \sqrt{\left|\frac{\frac{\partial E}{\partial P_{+-}}}{\frac{\partial E}{\partial P_{++}}}\right|} T_\rightarrow.
\end{equation}

Let us call the outgoing state after the split $\Psi^{(i)}_+$ and the ingoing state $\Psi^{(i)}_-$. We then can consider the first splitting a transformation of the wavefunction such that 
\begin{equation}
   \Psi_0=\psi_{-+} \rightarrow \left( \begin{array}{c} \Psi^{(i)}_+ \\ \Psi^{(i)}_- \end{array} \right)=\left( \begin{array}{c} R_{\leftarrow}\psi_{++} \\ T_{\leftarrow}\psi_{--} \end{array} \right).
\end{equation}
This splitting should preserve the probability current, for consistency. In the nonrelativistic, flat spacetime limit, the probability current $J$ satisfies the continuity equation
\begin{equation}
\frac{\partial}{\partial t}\left(|\psi|^2\right)+\frac{\partial}{\partial X} J=0
\label{eq:Cont}
\end{equation}
and is given by the usual quantum mechanics expression
\begin{equation}
\frac{1}{2im}\left(\psi^*\psi^\prime-\psi\psi^{*\prime}\right).
\label{eq:J0}
\end{equation}
Therefore, in this limit we have an input probability current of 
\begin{equation}
J_0 = |\Psi_0|^2 \frac{P_{-+}}{M_+}.
\label{eq:JInit}
\end{equation}

After first encountering the beam-splitter, the probability current (\ref{eq:JInit}) splits into reflected and transmitted components 
\begin{eqnarray}
|J_+^{(i)}| &=& \frac{1}{2iM_+}\left(\Psi_+^{(i)*}\left(\Psi_+^{(i)}\right)^\prime-\Psi_+^{(i)}\left(\Psi_+^{(i)}\right)^{*\prime}\right)\nonumber\\
&=& |J_0|\bar{R}_\leftarrow^2
\end{eqnarray}
and
\begin{eqnarray}
|J_-^{(i)}| &=& \frac{1}{2iM_-}\left(\Psi_-^{(i)*}\left(\Psi_-^{(i)}\right)^\prime-\Psi_-^{(i)}\left(\Psi_-{(i)}\right)^{*\prime}\right)\nonumber\\
&=& |J_0|\left(\frac{M_+ P_{--}}{M_- P_{-+}}\right)\bar{T}_\leftarrow^2 .
\end{eqnarray}
The splitting preserves probability current, as can be confirmed by observing that $\left| \frac{J_+^{(i)}}{J_0} \right|+\left| \frac{J_-^{(i)}}{J_0} \right|$ is unity. The terms $\left| \frac{J_+^{(i)}}{J_0} \right|$ and $\left| \frac{J_-^{(i)}}{J_0} \right|$ are usually called the reflection and transmission coefficients, respectively.

The second transformation propagates the modes along the interferometer arms, such that
\begin{equation}
\left( \begin{array}{c} \Psi^{(i)}_+ \\ \Psi^{(i)}_- \end{array} \right)
   \rightarrow \left( \begin{array}{c} \Psi^{(ii)}_+ \\ \Psi^{(ii)}_- \end{array} \right)=
	\left( \begin{array}{c} (E,_{++})^{-1/2}R_\leftarrow e^{i\Phi_{++}} \\ (E,_{--})^{-1/2}T_\leftarrow e^{i\Phi_{--}} \end{array} \right).
\end{equation}
For brevity, the notation $E,_{\pm\pm}$ was used to denote $\partial E / \partial P_{\pm\pm}$, and it is understood that we are evaluating these quantities at the outer walls of the interferometer. We have also introduced the quantities $\Phi_{\pm\pm}=\phi_{\pm\pm}-Et_{\pm\pm}$, for $\phi_{\pm\pm}=\int_{X_\delta}^{X_\pm}{dX P_{\pm\pm}}$, where $t_{++}$ and $t_{--}$ denote the travel times from the splitter to $X_+$ and $X_-$, respectively.

The modes then reflect off of the outer walls, as
\begin{equation}
\left( \begin{array}{c} \Psi^{(ii)}_+ \\ \Psi^{(ii)}_- \end{array} \right)
   \rightarrow \left( \begin{array}{c} \Psi^{(iii)}_+ \\ \Psi^{(iii)}_- \end{array} \right)=
	\left( \begin{array}{c} (E,_{-+})^{-1/2}R_\leftarrow e^{i\Phi_{++}}R^\rightarrow \\ (E,_{+-})^{-1/2}T_\leftarrow e^{i\Phi_{--}}R^\leftarrow \end{array} \right).
\end{equation}
The outer wall reflection amplitudes $\left(R^\rightarrow,R^\leftarrow\right)$ only depend on continuity of the wavefunction. To obtain the reflection amplitude from the left, for instance, consider the wavefunction 
\begin{displaymath}
   \Psi=\psi_{++} \rightarrow \left\{
     \begin{array}{lr}
       0 & : X > X_+ \\
       \left(\psi_{++}+R^{\rightarrow}\psi_{--}\right) & : X < X_+
     \end{array}
   \right.
\end{displaymath}
By applying wavefunction continuity at $X_+$, one immediately obtains $R^\rightarrow$. $R^\leftarrow$ can be similarly determined, and the results are
\begin{equation}
\bar{R}^\rightarrow=-e^{i\left(\phi_{++}+\phi_{-+}\right)}, \hspace{6pt} \bar{R}^\leftarrow=-e^{i\left(\phi_{+-}+\phi_{--}\right)},
\end{equation}
with help of the simplifying definitions
\begin{equation}
\bar{R}^\rightarrow \equiv \sqrt{\left|\frac{\frac{\partial E}{\partial P_{++}}}{\frac{\partial E}{\partial P_{-+}}}\right|} R^\rightarrow, \hspace{4pt} \bar{R}^\leftarrow \equiv \sqrt{\left|\frac{\frac{\partial E}{\partial P_{--}}}{\frac{\partial E}{\partial P_{+-}}}\right|} R^\leftarrow.
\end{equation}
The phases are defined such that $\phi_{\pm\mp}=\int_{X_\mp}^{X_\delta}{dX P_{\pm\mp}}$ (signs chosen together). We will refer to the modes after reflection from the outer walls as $\Psi^{(iii)}_\pm$.

Propagation along the arms back to the splitter then proceeds as
\begin{equation}
  \left(
     \begin{array}{c} \Psi^{(iii)}_+ \\ \Psi^{(iii)}_- \end{array} \right) 
	\rightarrow
		 \left(\begin{array}{c} \Psi^{(iv)}_+ \\ \Psi^{(iv)}_- \end{array} \right),
\end{equation}
for
\begin{equation}
\left(\begin{array}{c} \Psi^{(iv)}_+ \\ \Psi^{(iv)}_- \end{array} \right)=\left( \begin{array}{c} (E,_{-+})^{-1/2}R_\leftarrow e^{i\Phi_{++}}R^\rightarrow e^{i\Phi_{-+}} \\ (E,_{+-})^{-1/2}T_\leftarrow e^{i\Phi_{--}}R^\leftarrow e^{i\Phi_{+-}} \end{array} \right).
\end{equation}
In this expression, the quantities $E,_{-+}$ and $E,_{+-}$ are evaluated at the splitter, and we have used the definitions $\Phi_{\pm\mp}=\phi_{\pm\mp}-Et_{\pm\mp}$ (signs again chosen together). Here $t_{-+}$ and $t_{+-}$ denote the travel times from $X_+$ to the splitter and from $X_-$ to the splitter, respectively.

The second encounter with the splitter occurs as it did before, as
\begin{equation}
   \left(\begin{array}{c} \Psi^{(iv)}_+ \\ \Psi^{(iv)}_- \end{array} \right)\rightarrow  \left(\begin{array}{c} \Psi^{(v)}_+ \\ \Psi^{(v)}_- \end{array} \right)=
     \begin{pmatrix} \bar{R}_\leftarrow & \bar{T}_\rightarrow \\ \bar{T}_\leftarrow & \bar{R}_\rightarrow \end{pmatrix} \left( \begin{array}{c} \Psi^{(iv)}_+ \\ \Psi^{(iv)}_- \end{array} \right).
		\label{eq:FinalStates}
\end{equation}

At the order we are working at in $\hbar$, the derivatives of our final outputs satisfy
\begin{equation}
\frac{d}{dX}\Psi^{(v)}_\pm = iP_{\pm\pm}\Psi^{(v)}_\pm,
\end{equation}
and so the currents for our final outputs are given by
\begin{eqnarray}
J^{(v)}_\pm  &=& \frac{1}{2iM_\pm}\left(\Psi_\pm^{(v)*}\left(\Psi_\pm^{(v)}\right)^\prime-\Psi_\pm^{(v)}\left(\Psi_\pm^{(v)}\right)^{*\prime}\right)\nonumber\\
&=& \frac{P_{\pm\pm}}{M_\pm}\left|\Psi^{(v)}_\pm\right|^2.
\end{eqnarray} 

We then have enough information to calculate the final reflected and transmitted probability currents, which can be written 
\begin{equation}
|J_+^{(v)}| = |J_0|\left[1-4\bar{R}_\leftarrow^2\left(1-\bar{R}_\leftarrow^2\right)\sin^2{\varphi}\right]
\end{equation}
and
\begin{equation}
|J_-^{(v)}| = |J_0|4\bar{R}_\leftarrow^2\left(1-\bar{R}_\leftarrow^2\right)\sin^2{\varphi},
\end{equation}
where we have defined $\varphi=\phi_{++}+\phi_{-+}-\phi_{+-}-\phi_{--}$ and made use of the identity $\bar{R}_\leftarrow^2+\frac{M_+ P_{--}}{M_- P_{-+}}\bar{T}_\leftarrow^2=1$. The flat-spacetime interferometer thus manifestly conserves probability current in all regions of the parameter space.

One can now search for a nice region in the parameter space that cancels one of the outputs. First, we would like to avoid regions of the parameter space that don't describe splitting, i.e. complete initial reflection or transmission by the beam-splitter. We can accomplish this in a simple way by enforcing an equal splitting condition, $\bar{R}_\leftarrow^2=1/2$. This leads to compact expressions for the final reflection and transmission coefficients, given by
\begin{equation}
R_f \equiv \frac{|J_+^{(v)}|}{|J_0|} = \cos^2{\varphi}
\end{equation}
and
\begin{equation}
T_f \equiv \frac{|J_-^{(v)}|}{|J_0|}= \sin^2{\varphi},
\end{equation}
respectively. 

We should also make sure that our shell velocity doesn't approach the speed of light, since we are working in the nonrelativistic limit. For small shell speeds, given an outer mass $M_+$ and an initial speed $v_+$, the initial splitting will be equal provided the inner mass satisfies
\begin{equation}
M_- \approx M_+ \left[1+\left(6\sqrt{2}-8\right)v_+^2-\left(99\sqrt{2}-140\right)v_+^4\right].
\label{eq:Mminus}
\end{equation}
In the quantum context, the ``speed'' $v_+$ is defined such that $E=M_+ + \frac{1}{2} M_+ v_+^2$, for a WKB state with energy $E$.

If we denote the interferometer arm lengths by $L_\pm \equiv \pm (X_\pm-X_\delta)$, we can see from the form of the reflection and transmission coefficients that one of the outputs will be completely cancelled if
\begin{eqnarray}
\varphi &=& 2 L_+\sqrt{2M_+\left(E-M_+\right)}-2L_-\sqrt{2M_-\left(E-M_-\right)}\nonumber\\
&=& 2 L_+ M_+ v_+-2L_- M_- v_-\nonumber\\
&=& \frac{n\pi}{2},
\label{eq:cancel}
\end{eqnarray}
for $n\in \mathbb{Z}$. Thus, as the outer arm length is increased or decreased, the outputs are alternately cancelled out for each value of $n$ (odd values cancel the transmitted output, and even values cancel the reflected output), with partial interference for intermediate arm lengths that don't correspond to solutions of (\ref{eq:cancel}). This behaviour is a direct reflection of coherence in the flat spacetime interferometer.

\subsection{General Relativistic Picture}

The current framework was designed to facilitate the inclusion of general relativistic corrections. Several expressions become messier once one includes gravity, and some expressions fundamentally change in structure. For instance, the probability current given by (\ref{eq:J0}) is no longer conserved in systems with more general Hamiltonians. In fact, a probability current for an arbitrary Hamiltonian system has never been constructed; only special cases are known.

For our purposes, since we are working in the WKB regime, we may sometimes wish to approximate a general Hamiltonian $H(X,P)$ by the first three terms in a Taylor expansion in $P$, given by
\begin{equation}
H_w=H(X,0)+\left(\frac{\partial H}{\partial P}\right) P + \frac{1}{2}\left(\frac{\partial^2 H}{\partial P^2}\right) P^2,
\label{eq:HAp}
\end{equation}
with the $P$-derivatives evaluated at $P=0$. In the quantum theory, we can symmetrize the term linear in $P$ to enforce Hermiticity, i.e.
\begin{equation}
\left(\frac{\partial H}{\partial P}\right)P\rightarrow \frac{1}{2}\left(\hat{\left(\frac{\partial H}{\partial P}\right)}\hat{P}+\hat{P}\hat{\left(\frac{\partial H}{\partial P}\right)}\right),
\end{equation}
as well as ordering the quadratic term as
\begin{equation}
\left(\frac{\partial^2 H}{\partial P^2}\right) P^2 \rightarrow \hat{P}\hat{\left(\frac{\partial^2 H}{\partial P^2}\right)}\hat{P}.
\end{equation}

If we take this operator ordering of the approximate form (\ref{eq:HAp}) as an exact Hamiltonian, then we can find a probability current $J$ that satisfies the continuity equation (\ref{eq:Cont}), which we can express as
\begin{equation}
J = \left(\frac{\partial H}{\partial P}\right) |\Psi|^2 +\frac{1}{2i}\left(\frac{\partial^2 H}{\partial P^2}\right) \left(\Psi^*\Psi^\prime-\Psi\Psi^{*\prime}\right).
\label{eq:JG}
\end{equation} 
The $P$-derivatives in this expression are again evaluated at $P=0$, and for the special case of $\{\left(\frac{\partial H}{\partial P}\right)=0, \left(\frac{\partial^2 H}{\partial P^2}\right) = 1/m\}$, we are left with the nonrelativistic, flat spacetime limit described by (\ref{eq:J0}).

In the limit of large $X$, the WKB Hamiltonian for our shell system is given by
\begin{equation}
H_w\sim \left(\hat{M}-\frac{\hat{M}^2}{18 X}\right)-\frac{2}{3}\sqrt{\frac{2\hat{M}}{X}}P+\left(\frac{1}{2\hat{M}}+\frac{1}{3X}\right)P^2,
\label{eq:WKBHam}
\end{equation}
so the generalized probability current is given by
\begin{equation}
J\sim -\frac{2}{3}\sqrt{\frac{2\hat{M}}{X}}|\Psi|^2+\left(1+\frac{2\hat{M}}{3X}\right)J_{s},
\end{equation}
with $J_{s}$ being the standard (nonrelativistic) expression (\ref{eq:J0}) for the probability current. Note that although the functional form of $J_{s}$ with respect to $\Psi$ is the same as the nonrelativistic current (\ref{eq:J0}), in the above expression we are inserting the general relativistic WKB wavefunction $\Psi$.

Since our Schr\"{o}dinger equation now takes the asymptotic form
\begin{equation}
H_w \Psi = i \frac{\partial}{\partial t}\Psi,
\label{eq:Hw}
\end{equation}
taking the operator ordering mentioned above, we no longer have the simple reflection and transmission amplitudes obtained in the previous section. For instance, integrating (\ref{eq:Hw}) across $X_\delta$ yields
\begin{equation}
\left[\left(\frac{3}{2\hat{M}}+\frac{1}{X}\right)\Psi'\right]_\delta=i\sqrt{\frac{2}{X_\delta}}\left[\sqrt{\hat{M}}\right]_\delta \Psi(X_\delta).
\end{equation}
Here, $\left[\cdot\right]_\delta$ represents the jump of a quantity across $X_\delta$. 

\begin{figure*}
  \includegraphics[width=\textwidth,height=5.4cm]{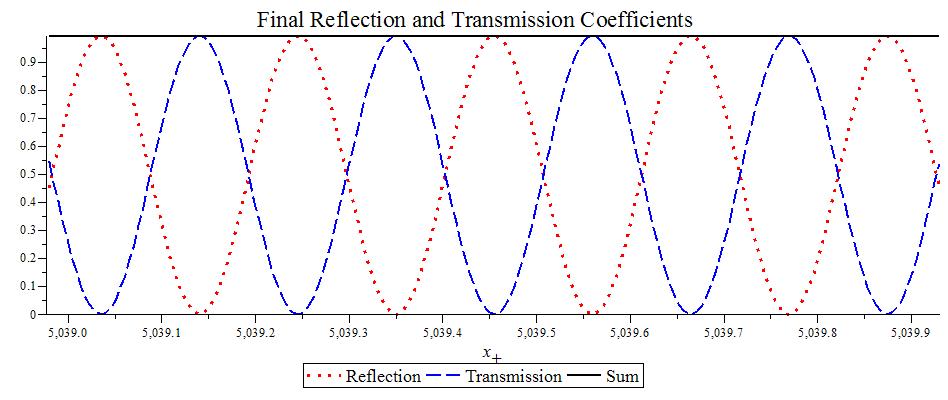}
  \caption{Sample interference pattern, for $M_+=15$, $v_+=0.003$, $X_-=5000$, $L_-=20$, and $M_-$ chosen to satisfy (\ref{eq:Mminus}), plotted against the outer mirror position, $X_+$.}
	\label{fig:Interference1B}
\end{figure*}

\begin{figure*}
  \includegraphics[width=\textwidth,height=5.4cm]{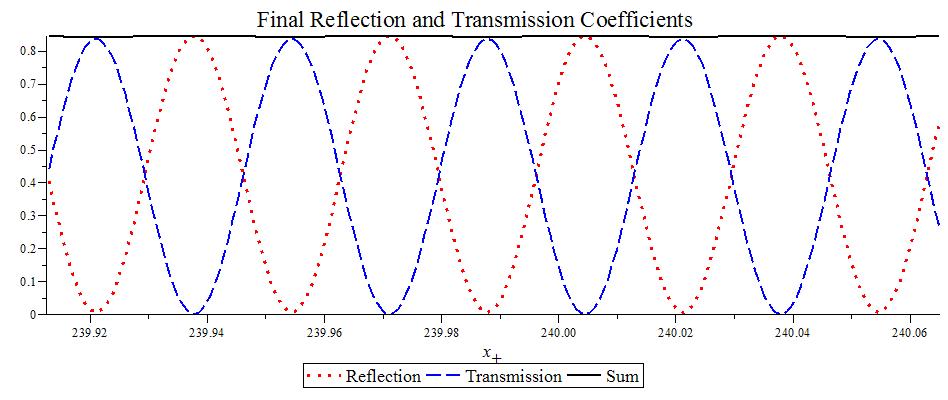}
  \caption{Sample interference pattern, for $M_+=15$, $v_+=0.01$, $X_-=200$, $L_-=20$, and $M_-$ chosen to satisfy (\ref{eq:Mminus}), plotted against the outer mirror position, $X_+$.}
	\label{fig:Interference2B}
\end{figure*}

To the order we are working at in $\hbar$, the new reflection and transmission amplitudes for scattering from the right are given by
\begin{widetext}
\begin{eqnarray}
\bar{R}_\leftarrow = \frac{\sqrt{\frac{2}{X_\delta}}\left[\sqrt{\hat{M}}\right]_\delta+\left(\frac{3}{2M_-}+\frac{1}{X_\delta}\right)P_{--}-\left(\frac{3}{2M_+}+\frac{1}{X_\delta}\right)P_{-+}}{-\sqrt{\frac{2}{X_\delta}}\left[\sqrt{\hat{M}}\right]_\delta-\left(\frac{3}{2M_-}+\frac{1}{X_\delta}\right)P_{--}+\left(\frac{3}{2M_+}+\frac{1}{X_\delta}\right)P_{++}}
\end{eqnarray}
and
\begin{eqnarray}
\bar{T}_\leftarrow =\frac{\left(\frac{3}{2M_+}-\frac{1}{X_\delta}\right)\left(P_{++}-P_{-+}\right)}{-\sqrt{\frac{2}{X_\delta}}\left[\sqrt{\hat{M}}\right]_\delta-\left(\frac{3}{2M_-}+\frac{1}{X_\delta}\right)P_{--}+\left(\frac{3}{2M_+}+\frac{1}{X_\delta}\right)P_{++}}.
\end{eqnarray}
Similarly, for scattering from the left we have
\begin{eqnarray}
\bar{R}_\rightarrow = \frac{\sqrt{\frac{2}{X_\delta}}\left[\sqrt{\hat{M}}\right]_\delta+\left(\frac{3}{2M_-}+\frac{1}{X_\delta}\right)P_{+-}-\left(\frac{3}{2M_+}+\frac{1}{X_\delta}\right)P_{++}}{-\sqrt{\frac{2}{X_\delta}}\left[\sqrt{\hat{M}}\right]_\delta-\left(\frac{3}{2M_-}+\frac{1}{X_\delta}\right)P_{--}+\left(\frac{3}{2M_+}+\frac{1}{X_\delta}\right)P_{++}}
\end{eqnarray}
and
\begin{eqnarray}
\bar{T}_\rightarrow = \frac{\left(\frac{3}{2M_-}-\frac{1}{X_\delta}\right)\left(P_{+-}-P_{--}\right)}{-\sqrt{\frac{2}{X_\delta}}\left[\sqrt{\hat{M}}\right]_\delta-\left(\frac{3}{2M_-}+\frac{1}{X_\delta}\right)P_{--}+\left(\frac{3}{2M_+}+\frac{1}{X_\delta}\right)P_{++}}.
\end{eqnarray}
\end{widetext}

Given the definition of probability current in this (more general) setting, we have 
\begin{equation}
J^{(v)}_\pm = \left(\left(1+\frac{2M_\pm}{3 X_\delta}\right)\frac{P_{\pm\pm}}{M_\pm}-\frac{2}{3}\sqrt{\frac{2M_\pm}{X_\delta}}\right)\left|\Psi^{(v)}_\pm\right|^2.
\end{equation} 
Just as in the flat spacetime limit, the final output states are given by (\ref{eq:FinalStates}), except that now the reflection/transmission amplitudes and the WKB phases are more complicated.

\begin{figure*}
  \includegraphics[width=\textwidth,height=5.9cm]{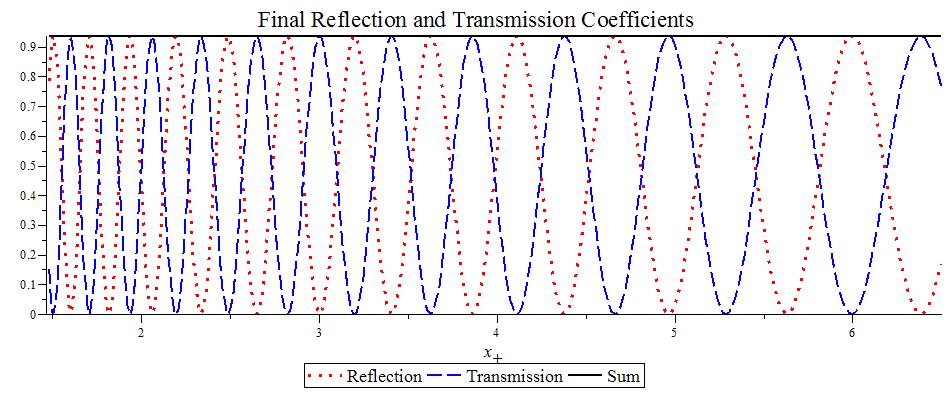}
  \caption{Sample interference pattern, for $M_+=0.05$, $v_+=0.0001$, $X_-\approx 1$, $L_- \approx 0.997$, and $M_-$ chosen to satisfy (\ref{eq:Mminus}), plotted against the outer mirror position, $X_+$.}
	\label{fig:Interference3B}
\end{figure*}

The initial current can be expressed as
\begin{equation}
J_i = \left(1-\frac{2M_+}{3P_{-+}}\sqrt{\frac{2M_+}{X_\delta}}+\frac{2M_+}{3X_\delta}\right)J_0,
\end{equation}
with $J_0$ being the nonrelativistic initial current (\ref{eq:JInit}), and so the final reflection and transmission coefficients $R_f \equiv \frac{|J_+^{(v)}|}{|J_i|}$ and $T_f \equiv \frac{|J_-^{(v)}|}{|J_i|}$ (respectively) are fully determined.

Another similarity to the flat spacetime limit is that the oscillatory part of the final reflection and transmission coefficients is defined by $\varphi=\phi_{++}+\phi_{-+}-\phi_{+-}-\phi_{--}$, with the various $\phi$ terms involving integrals of the general relativistic momentum (\ref{eq:Pc}). In the weak-field limit, the initial ingoing momentum is given by
\begin{eqnarray}
P_{-+}\sim -\sqrt{H^2-M_+^2}+\frac{2}{3}\sqrt{\frac{2H}{X}}H\nonumber\\
-\frac{\left(H^2-M_+^2/2\right)H}{\sqrt{H^2-M_+^2}X},
\end{eqnarray}
to second order in $1/\sqrt{X}$. Care should be taken with these approximations, however, because our probability current (\ref{eq:JG}) is exactly conserved only in the quadratic momentum limit, which for the shell system is defined by (\ref{eq:WKBHam}). Also, the WKB solutions only approximately satisfy the Schr\"{o}dinger equation. Because of this, in order to control the errors involved in the approximations we find it useful to consider the ``WKB momentum,'' which we define by solving (\ref{eq:WKBHam}) for $P$, and expanding to second order in $1/\sqrt{X}$. For the initial ingoing momentum, the WKB momentum takes the form
\begin{eqnarray}
P_{w-+}\sim-\sqrt{2M_+\left(H-M_+\right)}+\frac{2}{3}\sqrt{\frac{2M_+}{X}}M_+\nonumber\\
-\sqrt{\frac{2M_+}{\left(H-M_+\right)}}\frac{\left(7M_+-4H\right)}{12 X}.
\end{eqnarray}

To understand the interference pattern described by $R_f$ and $T_f$, let us consider what $\varphi$ looks like in the weak-field limit, for slow speeds ($v_\pm \rightarrow 0$):
\begin{eqnarray}
\varphi\hspace{7pt} \sim &2L_+ M_+ v_+-2L_- M_- v_-\nonumber\\
&+\frac{M_+^2}{v_+}\ln{\left(\frac{X_+}{X_\delta}\right)}-\frac{M_-^2}{v_-}\ln{\left(\frac{X_\delta}{X_-}\right)}.
\label{eq:cancelGR}
\end{eqnarray}
Let us further imagine that we vary the outer arm length $L_+$, while keeping all other parameters constant. If the phase condition (\ref{eq:cancel}) from flat spacetime still approximately holds, then the corresponding expression (\ref{eq:cancelGR}) in the weak-field limit tells us that successive values of $n$ (say, $n$ to $n+1$) are associated with outer arm length values $L_{+n}$ and $L_{+(n+1)}$. Subtracting $\varphi_n=n\pi/2$ from $\varphi_{n+1}=(n+1)\pi/2$ yields
\begin{equation}
\frac{\pi}{2}=2M_+ v_+ \Delta L_n +\frac{M_+^2}{v_+}\ln{\left(\frac{X_{+n}+\Delta L_n}{X_{+n}}\right)},
\end{equation}
with the definitions $X_{+n}=X_\delta+L_{+n}$ and $\Delta L_n =L_{+(n+1)}-L_{+n}$. The distance between nodes of the interference pattern, denoted by $\Delta L_n$, is somewhat less than the outer mirror radius $X_+$, for the cases we are interested in; thus, we can expand the logarithm and solve for $\Delta L_n$, which gives us
\begin{equation}
\Delta L_n \approx \frac{\pi}{4M_+\left(v_++\frac{M_+}{2v_+X_{+n}}\right)}.
\label{eq:nodes}
\end{equation}
This result shows that gravity causes the node spacing in the interference pattern to increase with increasing outer arm length. In the flat space limit (i.e. as $X_\pm\rightarrow \infty$), we obtain the equal node spacing $\Delta L_n =\pi/4M_+v_+$, for all $n\in\mathbb{Z}$.

One can see from Figures~\ref{fig:Interference1B} and \ref{fig:Interference2B} that as we go from the essentially flat limit ($X_\pm \rightarrow \infty$) to less than $10$ Schwarzschild radii, we can still alternately cancel the reflection and transmission coefficients, even though the approximations lead to a probability current that is not fully conserved (note that the sum of the final probability currents is about $15\%$ less than the initial current). We take this as a direct indication that coherence is fully present in the single-mode system even with general relativistic corrections taken into account.

It is not clear from Figures~\ref{fig:Interference1B} and \ref{fig:Interference2B}, but the node spacing is indeed changing as (\ref{eq:nodes}) suggests. The reason it is not visible from these plots is that the node spacing changes noticeably only over a range of many wavelengths. Under more extreme circumstances, as depicted in Figure~\ref{fig:Interference3B}, there are visible changes in node spacing, though this represents a situation that is of less physical interest, since the de Broglie wavelength of the shell is larger than the interferometer arms.

\section{Discussion}

There are two problems with taking this result of no loss of coherence as the definitive answer to whether or not gravity, by itself, could decohere a system. The first is that these single-mode states correspond in some sense to energy eigenstates; one might expect it is only a superposition of energies that leads to decoherence, since from the above analysis one can see that the time-dependence cancels out of the final expressions for output probabilities in the interferometer. As mentioned above, a study of how wave-packets behave in this model will be presented in a forthcoming paper \cite{GU14}. 

The second problem is that Penrose's intuition ties the loss of coherence to the inability to map one spacetime in any unique way onto a different spacetime. By our coordinate choice we have, in effect, chosen a unique way: two spacetime points are the same if they have the same coordinates. However, this is of course arbitrary and depends on the coordinate choice made. While the Painlev\'e-Gullstrand coordinates have many advantages, they are not the only possibile choice. A one-parameter family of generalized Painlev\'e-Gullstrand coordinates \cite{PainleveFamily} can also be used to perform analogous calculations to those above. Do all coordinate fixings produce the same maxima and minima in the interference pattern? The canonical momentum in the reduced system certainly depends on the coordinate choice, but one can show that a broad set of choices lead to the same classical action \cite{Menotti}. Still, it is unclear whether this is enough to ensure coordinate independence in the quantum setting. These issues will be examined in future work \cite{GQU14}.

While we do not report a result that demonstrates intrinsic decoherence due to gravity here, we have provided a model system for further analysis that could potentially lead to such a demonstration. This work can therefore be considered a first step towards a concrete derivation of Penrose's predictions within canonical quantum gravity. The main point argued here is that Penrose's initial arguments rely solely on the principles of QT and GR, and so we should thoroughly explore the possibility that his predicted decoherence could be shown to result solely from QT and GR before adding any assumptions about new physics.

It could also be argued that the reduced phase space approximation leads to an artificial form of time-evolution that is not entirely consistent with the ``timeless'' structure of canonical quantum gravity. For instance, the lack of a satisfactory interpretation of reduced phase space minisuperspace quantum cosmology was discussed in \cite{UnruhWald}. One might then be drawn to the conclusion that in the limited setting of our approximations, the evolution will necessarily be unitary (by construction), and we will escape Hawking's original arguments about pure states evolving into mixed states \cite{Hawking82} by virtue of our approximation scheme. 

Certainly, our simple model does not have the features often associated with nonunitary modifications to standard Hamiltonian evolution (such as the inclusion of microscopic wormhole interactions \cite{Ellis89}), but there is still reason to believe the evolution defined by (\ref{eq:Pc}) could in principle exhibit decoherence. For one thing, we have in some places used an approximate Hamiltonian (\ref{eq:WKBHam}) that is quadratic in momenta and strictly Hermitian, but it may not be possible to define a Hermitian Hamiltonian operator that exactly corresponds to the solution of (\ref{eq:Pc}) (which is transcendental). For another, even if one could solve (\ref{eq:Pc}), the resulting Hamiltonian would be non-polynomial in both the momenta and the coordinate $X$. This means that the time evolution of the wavefunction at $X$ is not described by a finite number of derivatives at $X$, and is thus nonlocal, in the sense that the evolution equation is equivalent to an integro-differential equation with finitely-many derivatives \cite{Sucher63}-\cite{Garbaczewski14}. While some systems can be nonlocal in this way and yet maintain coherence (such as in the case of relativistic particles in flat spacetime \cite{Sucher63}, \cite{Lammerzahl92}), in other such systems there can be unexpected behaviour such as ``nonlocally-induced randomness'' \cite{Garbaczewski13}, \cite{Garbaczewski14}, which would in our case be attributable to gravity. These studies are still in their infancy, so it remains an open question whether or not this type of nonlocal behaviour can be connected with gravitational intrinsic decoherence.   

Regardless of what the true theoretical mechanism is, the arguments for the existence of the decoherence effect studied here are compelling, and experimental investigations are already underway to test for signatures in micro-optomechanical systems \cite{PenroseBouwmeester}, \cite{PikovBouwm08}. There are many technical obstacles to overcome to minimize the effects of standard environmental decoherence (which obscures the desired behaviour), but there is hope that these types of experiments will bear fruit within the next decade \cite{Penrose14}. 


\section{Acknowledgements}

The authors would like to thank the Natural Sciences and Engineering Research Council of Canada (NSERC) and the Templeton Foundation (Grant No. JTF $36838$) for financial support, as well as Friedemann Queisser for helpful discussions.



\end{document}